\documentclass[aps,prl,reprint,superscriptaddress]{revtex4-2}
\usepackage[centertags]{amsmath}
\usepackage{amsfonts,amssymb,bm,graphicx,subfigure}
\usepackage[notrig]{physics}
\usepackage{orcidlink,hyperref}

\UseRawInputEncoding

\newcommand{\ud}{\mathrm{d}}

\begin{document}
\title{
The induced friction on a probe moving in a nonequilibrium medium }

\author{Ji-Hui Pei \orcidlink{0000-0002-3466-4791}}
\affiliation{Department of Physics and Astronomy, KU Leuven, 3000, Belgium} 
\affiliation{School of Physics, Peking University, Beijing, 100871, China}
\author{Christian Maes \orcidlink{0000-0002-0188-697X}}
\email{christian.maes@kuleuven.be}
\affiliation{Department of Physics and Astronomy, KU Leuven, 3000, Belgium}

\begin{abstract}
Using a powerful combination of projection-operator method and path-space response theory, we derive the fluctuation dynamics of a slow inertial probe coupled to a steady nonequilibrium medium under the assumption of time-scale separation.  
The nonequilibrium is realized  by external nongradient driving on the medium particles or by their (athermal) active self-propulsion. 
The resulting friction on the probe is an explicit time-correlation for medium observables and is decomposed into two terms, one entropic, proportional to the noise variance as in the Einstein relation for equilibrium media, and a frenetic term that can take both signs. 
As an illustration, we give the exact expressions for the linear friction coefficient and noise amplitude of a probe in a rotating run-and-tumble medium. 
We find a transition to absolute negative probe friction as the nonequilibrium medium exhibits sufficient and persistent rotational current. There, the run-away of the probe to high speeds realizes a nonequilibrium-induced acceleration.  Simulations show that its speed finally saturates, yielding a symmetric stationary probe-momentum distribution with two peaks.
\end{abstract}
\maketitle

\textit{Introduction.}
Obtaining reduced and effective descriptions is the bread and butter of statistical physics.  
One successful example is the understanding of hydrodynamics emerging from an underlying many-body dynamics, to characterize the long-wavelength and long-time behavior of fluids. 
Another example is the appearance of fluctuation dynamics such as Brownian motion \cite{Brown1828,Einstein1905,Smoluchowski1906,VanKampen1986,Duplantier2006} for a micrometer-sized colloidal tracer (or probe) suspended in a fluid from the atomistic picture of nature. 
More broadly, such Langevin dynamics appears over and over again in multifold applications and is central to many nonequilibrium studies \cite{Zwanzig2001,Kampen2011}.

The derivation of a reduced dynamics is a fascinating problem that requires technical and conceptual care.  
Part of that concerns the transfer of principles and properties from one level of description to a more coarse-grained level. 
One crucial result that has met with great success is the identification of the fluctuation-dissipation relation of the second kind (FDRII) \cite{Callen1951,Kubo1966,Chandler}, {\it aka} the Einstein relation, which relates the friction on a probe with the thermal noise.  
That connection is not {\it just there} to produce the (correct) equilibrium distribution; the deeper reason is the time-reversal invariance in equilibrium. 
On the technical side, besides exactly solvable models \cite{Zwanzig1973,Caldeira1983}, methods exist such as via Mori's or Zwanzig's projection-operators \cite{Zwanzig1961,Mori1965}, to derive reduced descriptions for general equilibrium media, 
extending to relativistic \cite{Dunkel2006}, astrophysical \cite{Chandrasekhar1949}, and quantum-statistical setups \cite{Nakajima1958,Breuer2009}.  
%That includes the projection-operator technique, pioneered in the work of Mori and Zwanzig \cite{Zwanzig1961,Mori1965}, which starts from the atomic description and Newtonian dynamics. 
%For stochastic medium, there is a new path-space approach, \cite{trajectories}, which works with the trajectory probability. 

%see also \cite{gen} for a discussion in the context of the generalized Langevin equation, and when adding nonconservative forces to act on the probe.

However, when dealing with a {\it steady nonequilibrium} medium coupled to a probe, often encountered in microrheology \cite{Cicuta2007,Xia2018,Waigh2005}, physics of Life \cite{Wu2000,Chen2007,Wirtz2009,Maggi2017,Mitterwallner2020,Seyforth2022}, and for self-propelled active fluids, \cite{Paul2022}, things complicate. 
Some recent studies on the generalized Langevin equation have even suggested that the FDRII is still valid \cite{Jung2021,Zhu2022,Zhu2023}.  
Yet, more detailed analyses on concrete models have contradicted these claims \cite{Sarracino2010,Jung2022,Netz2018,Doerries2021}. 
FDRII indeed requires the assumption that the nonequilibrium condition does not directly affect the media \cite{Harada2006}.
There are other case studies mainly focusing on fluids with steady flows \cite{Shea1996,SantamariaHolek2001,Holzer2010,Szamel2004,Krueger2009}, but the general structure of friction and noise has still been missing.

Up to now, the most satisfying result is given by the path-space response theory \cite{Maes2014,stef,act,Tanogami2022}, which works with trajectory probabilities. 
That path-space approach \cite{resp,Tanogami2022} not only shows the violation of FDRII but also produces measurable expressions for the friction coefficient. Yet, several strict assumptions have remained, and it was unclear what further approximations are allowed and useful. 
On the other hand, Zwanzig's nonlinear projection approach \cite{Zwanzig1961} is more formal and systematic and was applied very recently for nonequilibrium media  \cite{Solon2022}. 
Unfortunately, for nonequilibrium media, the resulting friction from that projection approach is not expressed in terms of measurable quantities, making it operationally more problematic.

In the present Letter, we derive the fluctuation dynamics for a slow inertial probe coupled to a nonequilibrium medium. 
We start with exposing the projection-operator technique in a general setup. 
Then, we apply nonequilibrium response theory to obtain a measurable expression for the friction. 
In the end, a general and extended FDRII appears, where a new friction-noise relation benchmarks the reduced dynamics for probes in a nonequilibrium medium.  
As illustrated with two examples (a run-and-tumble and a sheared medium), we see how the frenetic contribution \cite{frenesy}, may generate absolute negative friction, making the probe run away to high energies.

\begin{figure}
    \centering
    \includegraphics[width=0.4\textwidth]{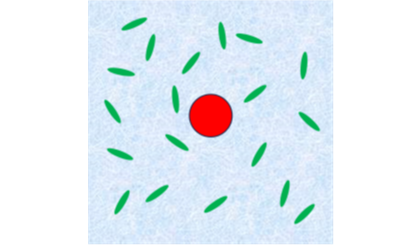}
    \caption{Sketch of a heavy probe (red ball) immersed in a nonequilibrium medium: 
    The green rods represent medium particles, out of equilibrium either by steady driving or self-propulsion, dissipating in the thermal environment (blue background). 
    }
    \label{levels}
\end{figure}

\textit{Setup. }
We consider a probe immersed in a medium consisting of many particles.  The medium is out of equilibrium due to either a nongradient steady driving or active self-propulsion. 
 The underlying equilibrium environment, in which the medium particles dissipate their energy, is described effectively in the (coarse-grained) dynamics of the medium; see Fig. \ref{levels}.  
 The probe is heavy, so its motion is much slower than the medium. 
For simplicity of presentation, we assume the probe interacts only with the medium particles. 
The degrees of freedom of the probe are represented by $\mathcal{A}$,
while the degrees of freedom of the medium are denoted by $\mathcal{B}$. 
We are interested in the reduced dynamics of the probe, {\it i.e.}, integrating out the medium.

Suppose the probe+medium dynamics can be described as a Markov process, with evolution equation $\dot \rho = L^\dagger \rho$
for the probability distribution $\rho(\mathcal{A},\mathcal{B};t)$ in terms of the forward generator $L^\dagger$, which is divided into two terms, 
$
    L^\dagger = L_0^\dagger +\epsilon L_1^\dagger 
$. 
The medium dynamics is represented here by $L_0^\dagger$, acting on $\mathcal{B}$ for fixed $\mathcal{A}$. 
%It contains the nonconservative effects from the external or active driving and the dissipation effect from the underlying environment (blue background in Fig. \ref{levels}). 
The generator $L_1^\dagger$ acting on $\mathcal{A}$ for fixed $\mathcal{B}$ is conservative. 
The small parameter $\epsilon$ represents the time-scale separation between the medium and the particle. 
We denote by $f_\mathcal{A}(\mathcal{B})$ the pinned (or Born-Oppenheimer) distribution, the stationary distribution of the medium dynamics for fixed $\mathcal{A}$, satisfying $L_0^\dagger f_\mathcal{A}=0$. 
For nonequilibrium, $f_\mathcal{A}(\mathcal{B})$ should be distinguished from the conditional stationary distribution $\rho_\text{st}(\mathcal{B}|\mathcal{A})=\rho_\text{st}(\mathcal{A},\mathcal{B})/\int\mathrm{d}\mathcal{B} \rho_\mathrm{stat}(\mathcal{A},\mathcal{B})$.

\textit{Projection-operator method.}
A satisfactory derivation of the projection-operator technique for a steady nonequilibrium medium is given in \cite{Solon2022}. 
However, for a shaped probe, \cite{Baek2018,Granek2022}, for sheared inhomogeneous media \cite{SantamariaHolek2001,Holzer2010}, or for two probes in one medium \cite{Dzubiella2003,Hayashi2006},
a derivation of the reduced dynamics is still lacking. 
In the Supplementary Material (SM) \cite{SM}, we provide such a derivation, serving as a generalization of previous results. 
The only approximation is the time-scale separation,  characterized by the small constant $\epsilon $. 
We track terms to order $\epsilon^2$ and focus on the behavior at the time scale $t\sim O(\epsilon^{-2})$. 
In that limit, the reduced dynamics is automatically Markovian \cite{SM}. 

\textit{General structure.}
The resulting dynamics for the probe 
can be expressed by a formal Fokker-Planck equation for the reduced distribution $\tilde \rho(\mathcal{A},t)=\int\ud \mathcal{B}\, \rho(\mathcal{A},\mathcal{B};t)$: 
\begin{align}
    \pdv{t}\tilde \rho(\mathcal{A},t)&=-\sum_i\pdv{\mathcal{A}_i}\left[K_i(\mathcal{A}) \tilde \rho(\mathcal{A},t)\right]\notag\\
    &\hspace{-1.5cm} -\sum_{i}\pdv{\mathcal{A}_i}\left[\mu_i(\mathcal{A}) \tilde \rho(\mathcal{A},t)\right] %\\
    %&
    -\sum_{i}\pdv{\mathcal{A}_i}[\sum_j\pdv{D_{ij}(\mathcal{A})}{\mathcal{A}_j} \tilde \rho(\mathcal{A},t)]\notag\\
    &+\sum_{i,j}\pdv{\mathcal{A}_i}\pdv{\mathcal{A}_j}\left[D_{ij}(\mathcal{A}) \tilde \rho(\mathcal{A},t)\right] .\label{FPE} 
    \end{align}
The summations are over all the slow degrees of freedom; {\it e.g.}, $\mathcal{A}_i $ refers to the position and momentum of the probe, and
\begin{align}
    &K_i     = \big\langle L \mathcal{A}_i\big\rangle_\mathcal{A}^\text{bo} ,\\
\label{fro}
    &\mu_i =-\int_0^\infty \ud s \sum_j\Big\langle L \mathcal{A}_i(s) ; L\mathcal{A}_j\,\frac{\partial }{\partial \mathcal{A}_j}\log f_\mathcal{A}\Big\rangle_\mathcal{A}^\text{bo},\\
\label{dif}
    &D_{ij} =\int_0^\infty \ud s\,\big\langle  L\mathcal{A}_i(s) ; L\mathcal{A}_j(0)\big\rangle_\mathcal{A}^\text{bo} .
\end{align}
Here, $\langle g\rangle_\mathcal{A}^\text{bo}=\int \ud \mathcal{B} f_\mathcal{A}(\mathcal{B})g(\mathcal{A},\mathcal{B})$ represents the average of $g$ over the pinned distribution $f_\mathcal{A}$ of the medium, and $\langle \cdot\,;\,\cdot\rangle_\mathcal{A}^\text{bo}$ is the time-dependent connected correlation (covariance) in the pinned dynamics (keeping $\mathcal{A}$ fixed). 
$K_i$, of order $\epsilon$, is the induced quasistatic force on the probe.
$\mu_i$ is a drift term related to dissipation, and the noise term is $D_{ij}$, both of order $\epsilon^2$. 
The symmetrization  $(D_{ij}+D_{ji})/2$ is a positive-definite matrix. 
The third term in \eqref{FPE} is the drift originating from possibly inhomogeneous media.

Simplifications occur when the probe is a point particle of mass $m$ with position $q$ and momentum $p$, and the coupling between the medium $\eta$ and the probe only involves the position $q$ via some potential $U(q,\eta)$.  (We replace $\mathcal{B}$ with $\eta$ to denote all the medium degrees of freedom in the following.)
Then, the pinned distribution $f_q$ only depends on the probe position. We have $Lq = p/m$, and $Lp=F(q,\eta)=-\partial_qU$ is the total force on the probe. 
In that case, the probe dynamics is described by $m\dot{q} = p$ and (in It\^o's convention)
\begin{equation}\label{new}
\dot{p} = \langle F(q,\eta)\rangle^\text{bo}_q - \gamma(q) \frac{p}{m} + \sqrt{2B(q)}\xi(t)  ,
\end{equation}
where $\langle F(q,\eta)\rangle^\text{bo}_q$ is the force on the probe by the medium in the quasistatic approximation, 
and the linear friction coefficient $\gamma$ (extracted from the drift term in \eqref{fro}) and noise coefficient $B$ (amplitude of noise variance) are 
\begin{align}
&\gamma=\int_0^\infty \ud s \,\langle F(q,\eta(s));\frac{\partial}{\partial q}\log f_q\rangle_q^\text{bo}, \label{gamma}\\
 &   B=\int_0^\infty \ud s\, \langle  F(q,\eta(s)); F(q,\eta(0))\rangle_q^\text{bo} .\label{D}
\end{align}
%$B$ is related to the diffusion coefficient $D$ by $B=D\gamma^2$. 
%The average $\langle\cdot\rangle_q^\text{bo}$ over the pinned distribution for medium only depends on the probe position $q$ now. 

In the special case of an equilibrium medium at inverse temperature $\beta$,  $f_q(\eta)\propto\exp(-\beta U(q,\eta))$, and $\partial_q \log f_q = \beta F(q,\eta)$
can be substituted in \eqref{gamma}, resulting in the standard FDRII, $\gamma=\beta B$.
However, for a nonequilibrium medium, there is no canonical form for the pinned distribution $f_q(\eta)$, and hence Eq. \eqref{gamma} is not directly measurable.  That is remedied in the following.

\textit{Nonequilibrium response theory. }
To accommodate an observable version of the friction, we need to replace the  $\partial_q\log f_q$ in \eqref{gamma} with more operationally accessible quantities.  That can be achieved by the use of nonequilibrium response theory, developed from a path-space approach 
\cite{resp,upd,over,Baiesi2009}.

We define $\gamma(s)=\langle F(q,\eta(s)) \,\partial_q\log f_q\rangle_q^\text{bo}$, where $\gamma=\int_0^\infty\ud s\gamma(s)$, and the observable $F(q,\eta(s))$ must be evaluated at time $s\geq 0$ according to the pinned dynamics for fixed $q$.  
Since $\partial_q\log f_q=\partial_q f_q/f_q$, we have (to linear order in $\Delta q$) 
\begin{equation}\label{li}
    \gamma(s)\Delta q =\int\ud \eta\, e^{sL_0}F(q;\eta) \,[f_{q+\Delta q}(\eta) -  f_{q}(\eta)] ,
\end{equation}
where we used the backward generator $L_0$ of the medium (depending on fixed $q$). 
In order to make \eqref{li} and hence \eqref{gamma} operational, we introduce a quenched process for the medium, where the probe position keeps $q+\Delta q$ from the infinite past until time $t=0$; at $t=0$, the probe position is suddenly and forever changed to $q$. 
As a result, the medium has distribution $f_{q+\Delta q}$ at $t=0$, and then undergoes a relaxation process with fixed $q$.  
With $\int\ud \eta\, e^{sL_0}F(q;\eta)f_{q+\Delta q}(\eta) = \langle F(q;\eta(s))\rangle^*_{q+\Delta q\rightarrow q}$  for the average of $F$ in this perturbed process, the above equation \eqref{li} is expressed as 
\begin{equation}\label{lin}
\gamma(s)\Delta q  = \langle F(q;\eta(s))\rangle^*_{q+\Delta q\rightarrow q} - \langle F(q;\eta(s))\rangle_{q}^\text{bo} .
\end{equation}
That difference between the average in the perturbed dynamics and the pinned dynamics can be handled with the techniques of nonequilibrium response theory \cite{resp,upd}. 

To be concrete, we take the common case where the medium consists either of driven overdamped particles or of discrete states following a Markov jump process, in contact with an environment at inverse temperature $\beta $ in which heat is dissipated.  Applying linear response theory (see SM \cite{SM}), we obtain  
\begin{equation}\label{efr}
    \begin{aligned}
     \gamma(s)     =&\frac{\beta}2 [\langle F(q,\eta(s));F(q,\eta(0))\rangle_q^\text{bo}\\
     &-\int_s^\infty\ud u \langle L_0F(q,\eta(0))\;F(q,\eta(u))\rangle_q^\text{bo}].
    \end{aligned}
\end{equation}
All observables in the average are explicit which renders the expression for the friction operationally useful. Other but similar expressions apply for other media, such as underdamped or multi-temperature media \cite{under}. 

The above formula generalizes and clarifies previous work in \cite{Maes2014,stef}. 
In those studies, similar formulas are derived relying solely on nonequilibrium response theory, with some of the approximations remaining less clear. 
More recently, Tanogami \cite{Tanogami2022} employed singular perturbation theory to justify \eqref{efr} and applied it for a medium subject to potential switching \cite{Wang2016}.
In the present study, the validity of \eqref{efr} becomes firmly established, as it is a reformulation of \eqref{gamma}, which itself is systematically derived using the projection-operator method.

%Below, we give an example of an active bath as well.  
%\textbf{to delete??:However, when the medium dynamics is essentially nonMarkovian, there does not appear an immediate solution except for Gaussian dynamics \cite{gle,gle1,gle2}.  The latter still includes active and thermal Ornstein-Uhlenbeck medium particles, \cite{act}.  If the medium Born-Oppenheimer distribution is known, we can directly apply \eqref{new}, which is useful in some cases of active or resetting medium dynamics, \cite{res}.}\\
%In summary, the induced fluctuation dynamics \eqref{new} can be written as 
%\begin{equation}\label{newc}
%\dot{p} = F(q) - \gamma \frac{p}{} + \sqrt{2D}\,\xi(t) 
%\end{equation}

The friction $\gamma = \int_0^\infty \gamma(s)\ud s$ in the form \eqref{efr} and the noise amplitude in \eqref{D} imply the decomposition 
\begin{equation}\label{me}
\begin{aligned}
\gamma &=\gamma_\text{ent} + \gamma_\text{fren},\\
\quad B &= 2T\gamma_\text{ent} = \int_0^\infty\ud s\,\langle Lp(s);Lp(0)\rangle_q^\text{bo} ,
\end{aligned}
\end{equation}
with the second equality giving all that remains of the Einstein relation.  For an equilibrium medium, $\gamma_\text{fren} = \gamma_\text{ent}$ and the standard FDRII, $B = T \gamma$, is recovered.  Such relations extend beyond the case of nonequilibrium media in contact with a thermal environment as we show in the example of run-and-tumble particles below.  Indeed, that decomposition of the friction $\gamma$ follows the general scheme of entropic, respectively, frenetic contributions, \cite{frenesy}.  Both are explicit in terms of measurable quantities and time-correlation functions, as seen in \eqref{efr}, \cite{resp}.

%More explicitly, for a medium in equilibrium $L_0$ in the second term of \eqref{efr} is symmetric and 
%\[
 %\langle Lp(0)\,L_0Lp(u)\rangle_q^\text{bo} = \frac{\ud}{\ud u}\langle Lp(0)\;Lp(u)\rangle_q^\text{bo}
%\]

%%We can still rewrite it to make the difference with equilibrium more visible:
%\begin{equation}\label{efr}
%\gamma = \beta \langle Lp(s);Lp(0)\rangle_q^\text{bo} + \left[..........   %\right]
%\end{equation}
%where the second term is zero for an equilibrium medium. 

\textit{Emergence of absolute negative friction. }
A medium of run-and-tumble particles can be viewed as modeling (strongly) confined bacterial motion, at least in the low-density limit where vortex reversal is easier \cite{Wioland2013,Nishiguchi2024}.  
In the first example, we consider the medium consisting of $N$ independent particles with positions $x_t^i$ in overdamped run-and-tumble motion on a ring with unit radius. 
%The medium particles are propelled by a constant speed $v$, and the direction of the propulsion flips at rate $\alpha$. 
They are propelled by constant speed $v$ while the direction of propulsion flips at rate $\alpha$.  Medium particles attract a probe $(p_t,q_t)$ with large mass $m$ (also on the ring)  via a force $F_m$ that has constant magnitude and is always directed along the shorter arc between the medium particle and the probe, 
\begin{equation}\label{rtp}
\begin{aligned}
    &\dot x_t^i/\chi  = v\sigma_t^i + F_m(x_t^i-q_t),\qquad \sigma_t^i\overset{\alpha}{\rightarrow} -\sigma_t^i \\
    &\dot p_t = - \sum_i F_m(x_t^i-q_t), \qquad m\dot q_t= p_t ,
    \end{aligned}
\end{equation}
with $\chi$ the mobility of the medium particles, which will be set to $1$ in the following. 
We restrict the positions $x,q$ to the range $(-\pi,+\pi]$ and apply the periodic boundary condition. 
For $k>0$ we put,
\begin{equation}\label{ifF}
    F_m(x-q)=\begin{cases}
        -k , &x-q\in (0,\pi)\\
        k , &x-q\in (-\pi,0) \\ 0 , &x=0, \pi 
    \end{cases} .
\end{equation}
We take $v>k$ so that the medium particles can turn around the ring.\\

\begin{figure}
    \centering
\includegraphics[width=0.35\textwidth]{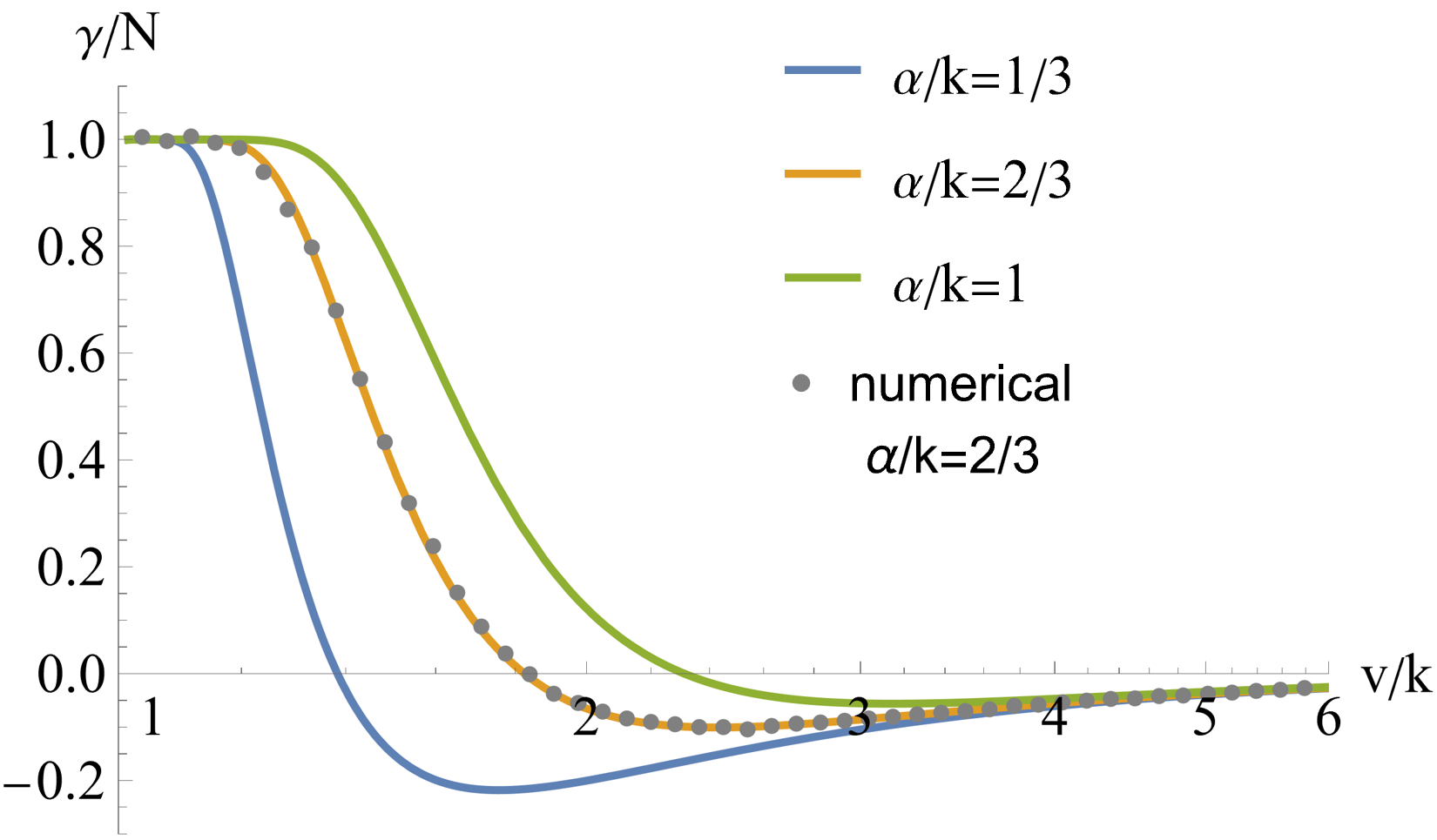}
    \caption{The friction coefficient at low probe speed for the run-and-tumble example \eqref{rtp}: Solid lines represent the analytic result as a function of $v/k$ for fixed $\alpha/k=1/3, 2/3,1$. 
    The friction coefficient from the simulation of the composite system with $\alpha/k=2/3$, $m=400$, $N=20$ is shown in gray dots. 
    }\label{fig:friction}
\end{figure}

For that example, we can compute all coefficients in  \eqref{new}; see SM \cite{SM}.  
%Refer to the supplementary materials for the derivations and the analytical expressions for the friction coefficient. 
Since medium particles are independent, the quasistatic force, the friction, and the (square root) noise amplitude $B$ are proportional to $N$. 
In the following, all quantities are per single medium particle. 
The quasistatic force $\langle F \rangle =0$ by rotation invariance. The linear friction coefficient $\gamma$, whose explicit expression is given in SM \cite{SM}, only depends on $v/k,\alpha/k$.
In Fig. \ref{fig:friction}, $\gamma$ is plotted as a function of $v/k$ for different $\alpha/k$. 
For small $v/k$, the friction coefficient remains close to $1$.
As $v/k$ increases, the friction coefficient decreases rapidly to negative values and reaches a minimum, which is deeper for smaller $\alpha/k$, and then increases to vanish when $v/k\rightarrow \infty$. 
For the limit of $\alpha\downarrow 0$, $\gamma\rightarrow -(k/v)^2$, and the negativity of $\gamma$ becomes most significant.

The friction decomposition \eqref{me} gives the entropic part $2\gamma_\text{ent} =  \beta_\text{eff}\,B$ for effective inverse temperature 
$
\beta_\text{eff}=2\alpha/( v^2-k^2)
$
where the square root noise amplitude $B$ is given in SM \cite{SM}.  We also find $\gamma_\text{ent} - \gamma_\text{fren}= \left((v/k)^2 - 1\right)^{-1}$ which goes to zero in the passive limit $v\uparrow \infty$ keeping $2\alpha/v^2$ constant.

\begin{figure}
    \centering
        \includegraphics[width=0.96\linewidth]{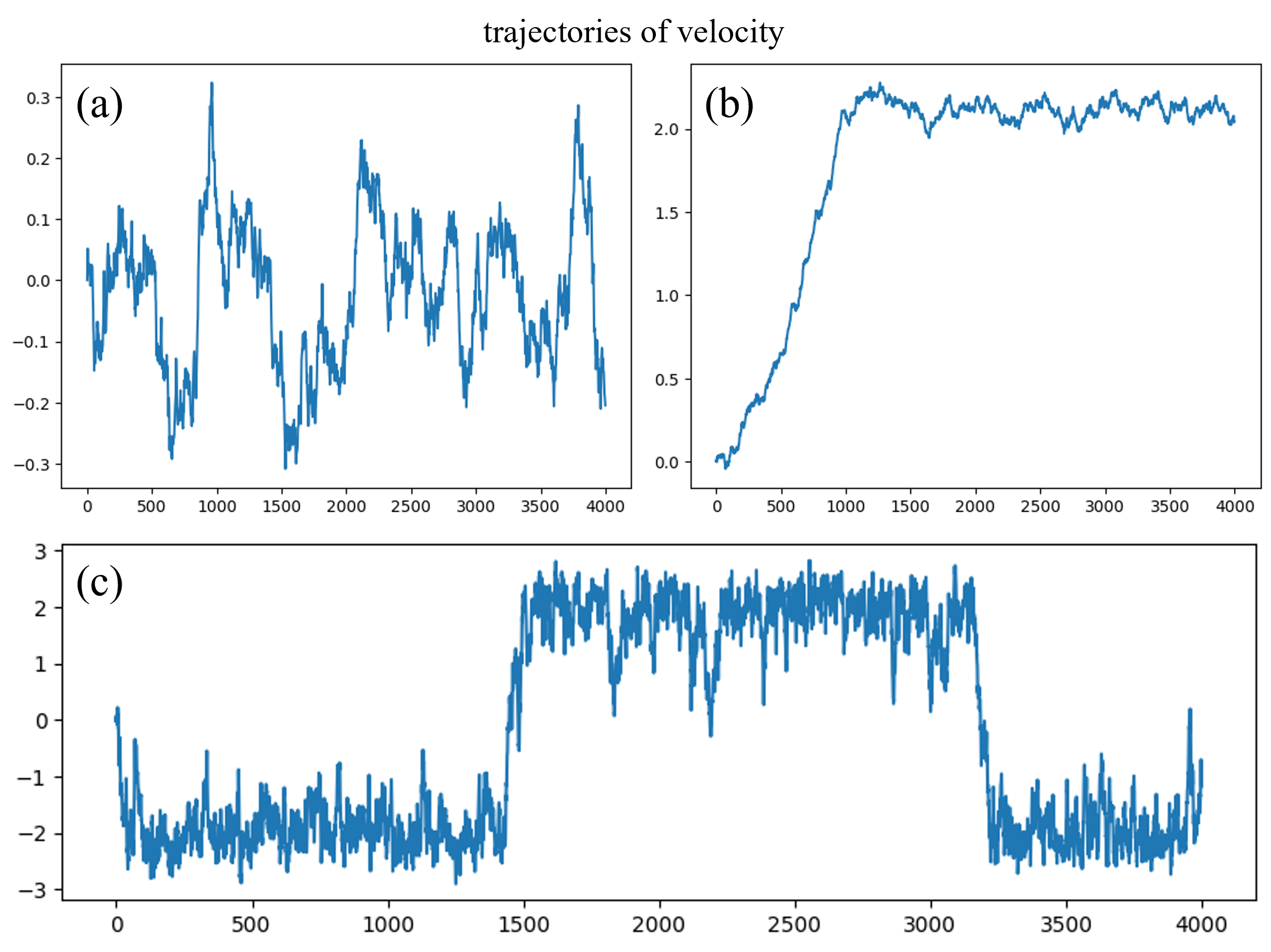}
    \caption{Typical velocities as function of time from the simulation of \eqref{rtp} for $N=20$, $k=1$, $\alpha=1$: with  parameters (a) $m=400$, $v=2$, (b) $m=400$,  $v=3$, (c) $m=16$, $v=3$.  }
    \label{fig:trajectory}
\end{figure}

The above analytic results regarding the friction coefficient (per medium particle) and the transition between positive and negative friction are confirmed by simulation for small probe speed \footnote{In the simulation, the friction coefficient is obtained by adding a positive external friction to make sure the overall friction is positive and by exerting a constant driving force: we  then measure the steady velocity of the probe by sampling trajectories}; see Fig. \ref{fig:friction}. 
We took a medium consisting of $N=20$ independent run-and-tumble particles, and we generate the trajectories of the composite system \eqref{rtp}, without any approximation. Furthermore, velocity trajectories are plotted in
Fig. \ref{fig:trajectory}. 
In the positive friction regime, the motion of the probe is just an underdamped Brownian motion. 
Yet, in the negative friction regime, the dynamics becomes very different: the probe spontaneously speeds up, reaching either a positive (counterclockwise) velocity or negative (clockwise) velocity with equal probability. 
The probe speed saturates at a specific value $v_s$, comparable to the propulsion speed $v$ of the RTP-particles. 
For small enough probe mass $m$, transitions between {velocity} $v_s$ and $-v_s$ are observed in the simulation, but not for large $m$.   We emphasize that the derived reduced dynamics is able to determine the transition point to the negative friction regime but does not explore the regime of high probe-speeds, where the time-scale separation is violated.
Simulations, as indicated by Fig.~\ref{fig:trajectory}, reveal that the stationary velocity distribution of the probe is bimodal with two sharp peaks. 

The above example clearly shows the possible emergence of absolute negative friction. 
That the friction contributed by an active medium can be partially negative is also reported in Ref. \cite{Granek2022}, but the overall friction is still positive, and the nonequilibrium-induced acceleration is not observed there. A related negative drag is studied in \cite{Kim2024}. 
On the other hand, for a harmonic interaction between 1-d (infinite boundary instead of periodic boundary) run-and-tumble particles  and a probe, the friction does not show any negativity \cite{ion2023}. 
A possible cause of the negative friction for \eqref{ifF} on the ring is the persistent rotational current in the medium for $v>k$.  Unsurprisingly then, the emergence of negative friction can also be observed for a thermal two-dimensional medium that is persistently sheared, as we show next.

Consider a two-dimensional medium with independent 2d overdamped Brownian particles with positions $(x^i,y^i)$ confined in a harmonic well (spring constant $k$) and driven by a rotational force (shear parameter $a \geq 0$). 
The probe with state $(p,q)$ is a one-dimensional heavy object.  The
$q$ and $x$ are coupled by a harmonic interaction with amplitude $b$.
The equations of motion are 
\begin{equation}
    \begin{aligned}
    \dot x&= -k x- a y -b(x-q)+\sqrt{2T}\xi_x\\
    \dot y&= a x -k y +\sqrt{2T}\xi_y\\
    \dot q&=p/m,\qquad
    \dot p= -b(q-x) .
    \end{aligned}
\end{equation}
where the thermal environment delivers temperature $T$ as amplitude for the standard white noises $\xi_x,\xi_y$ on each medium particle. 
The mobility of the medium particle is again set to $1$. 
The medium process (with fixed $q,p$) violates detailed balance for $a\neq 0$.

By linearity, all averages and time-dependent correlations in the general scheme can be easily calculated.  Integrating out the sheared medium, we find the induced quasistatic force $\ev{F}$, the friction coefficient $\gamma$, and the noise coefficient $B$ as follows, 
\begin{equation}
\begin{aligned}
    &\ev{F} = -\frac{b(k^2+a^2)}{a^2+bk+k^2} q, \quad
    \gamma=\frac{b^2(k^2-a^2)}{(a^2+bk+k^2)^2} ,\\
    &B=\frac{b^2(k^2+a^2)}{(a^2+bk+k^2)^2}\,T,\; \gamma_\text{fren} = \frac{b^2(k^2-3a^2)}{2(a^2+bk+k^2)^2}  .
\end{aligned}
\end{equation}
As before, all refer to quantities per single medium particle. 
In the equilibrium case $a=0$, we recover the Einstein relation $B= T \gamma$.  For $a\leq k$, the entropic contribution to the friction still dominates.
The transition occurs at $a=k$ where the friction vanishes, and the 
probe dynamics is effectively at infinite temperature $ \propto B/\gamma \rightarrow \infty$. For $a>k$, absolute negative friction is obtained, $\gamma<0$. 
The derived reduced dynamics for this harmonic model is valid for arbitrary velocity, meaning that the probe will be accelerated forever, which is different from the previous example. 

%We expect that diffusive processes with negative friction do not last forever. 
%The probe will move faster and faster and time-scale separation finally breaks down which is the key condition for our derivation \footnote{ There is no conflict with our assumption of having the probe time $t\sim O(\epsilon^{-2})$ because the friction is $O(\epsilon^2)$ }. 

\textit{Conclusion. }
We have derived the fluctuation dynamics of a probe after integrating out a fast nonequilibrium medium, by combining an extended projection-operator method with  path-space response theory.
The result is rather general. 
Even local detailed balance, \cite{local}, is not needed, implying a huge convenience for applications to complex dynamics, such as active media. 
Also for nonequilibrium media, with Zwanzig's projection method, the approximation of time-scale separation is easily tracked, and there is no need for weak coupling or for the probe to remain around a fixed position, as required in previous works \cite{stef,Maes2014}. 
On the other hand, there is the inaccessibility of the pinned distribution in \eqref{gamma} and, most importantly, that gets remedied by nonequilibrium response theory, leading to expressions such as \eqref{efr} and giving rise to the benchmarking relations \eqref{me}.

Friction is ubiquitous in any open system but may be modified by various means and for multiple purposes; see {\it e.g.}, \cite{fric,fric2}.  
The present study indicates it is possible to alter or even remove and reverse the effect of friction by appropriate driving of the medium to which the probe is coupled. 
%Two examples have shown how a persistent rotational current of the medium can generate absolute negative friction. 
The absolute negative friction phenomenon is of course a purely nonequilibrium effect: the probe absorbs energy from the medium, similar to but different from the scenario for stochastic or Fermi acceleration, \cite{Fermi1949,Sturrock1966}. \\

\begin{acknowledgments}
\textit{Acknowledgments.} This work got support from the China Scholarship Council, No. 202306010398. 
\end{acknowledgments}

\bibliography{reduced.bib}

\end{document}

\title{{\Large Supplementary Material}:
The induced friction on a probe moving in a nonequilibrium medium }
%, diffusive systems and deterministic dynamics 
% Wald entropy enters statistical mechanics

\author{Ji-Hui Pei \orcidlink{0000-0002-3466-4791}}
\affiliation{Department of Physics and Astronomy, KU Leuven, 3000, Belgium} 
\affiliation{School of Physics, Peking University, Beijing, 100871, China}
\author{Christian Maes \orcidlink{0000-0002-0188-697X}}
\affiliation{Department of Physics and Astronomy, KU Leuven, 3000, Belgium} 
%\date{\today}
\maketitle

\section{Derivation of the induced Fokker-Planck equation}\label{cal}
In this section, we derive the Fokker-Planck equation (Eq.~1 in the main text) for the reduced probe dynamics, using the projection-operator technique. 
This derivation is under a general setup of integrating out fast variables and can be regarded
as a generalization of \cite{Solon2022}. 

We use $\mathcal{A}$ to denote the probe degrees of freedom and $\mathcal{B}$ for the medium degrees of freedom that are waiting to be integrated out. 
The evolution of the distribution of the composite system, $\rho(\mathcal{A},\mathcal{B};t)$ is 
\begin{equation}\label{eom}
    \pdv{t} \rho = L^\dagger \rho .
\end{equation}
The forward generator is $L^\dagger= L_0^\dagger+\epsilon L_1^\dagger$ of the original Markov dynamics,  where the time-scale separation between the medium and the probe dynamics is indicated by small constant $\epsilon$. 

We introduce the projection operator $P^\dagger$ acting on a joint distribution function $\rho(\mathcal{A},\mathcal{B})$, 
\begin{equation}
    P^\dagger \rho\,(\mathcal{A},\mathcal{B}) = f_\mathcal{A}(\mathcal{B})\int \ud \mathcal{B}^\prime \rho(\mathcal{A},\mathcal{B}^\prime) .
\end{equation}
That projection integrates out the medium and replaces the medium distribution with the pinned distribution $f_\mathcal{A}(\mathcal{B})$, the stationary distribution of the medium dynamics with $\mathcal{A}$ fixed. 
The conjugate projection operator $P$ is defined on a function $g(\mathcal{A},\mathcal{B})$ by: 
\begin{equation}\label{bov}
    (P g)(\mathcal{A}) = \int \ud \mathcal{B} g(\mathcal{A},\mathcal{B}) f_\mathcal{A}(\mathcal{B}) =: \langle g\rangle_\mathcal{A}^\text{bo} ,
\end{equation}
where  $\langle g\rangle_\mathcal{A}^\text{bo}$ is the average of $g$ over the pinned distribution $f_\mathcal{A}$ of the medium.
Since $f_\mathcal{A}(\mathcal{B})$ is an invariant measure of $L_0^\dagger$, it follows that
\begin{equation}
    P^\dagger L_0^\dagger=L_0^\dagger P^\dagger=0 .
\end{equation}

We denote $Q^\dagger = 1-P^\dagger$. 
Projecting \eqref{eom} with $P^\dagger $ and $Q^\dagger $ from the left side and noticing that $P^\dagger$ and time-derivative commute, we have the following two equations, 
\begin{equation}
    \begin{aligned}
    \pdv{t}P^\dagger \rho = P^\dagger L^\dagger P^\dagger \rho + P^\dagger L^\dagger Q^\dagger \rho \\
    \pdv{t} Q^\dagger \rho = Q^\dagger L^\dagger P^\dagger \rho+ Q^\dagger L^\dagger Q^\dagger \rho
    \end{aligned}
\end{equation}
Solving $Q^\dagger\rho$ in the second equation and plugging it in the first equation, we obtain an exact equation for the projected distribution $P^{\dagger} \rho(t)$, 
\begin{equation}\label{exact}
%\begin{aligned}
   \frac{\partial}{\partial t} P^{\dagger} \rho(t)=P^{\dagger} L^\dagger P^{\dagger} \rho(t)+P^{\dagger} L^\dagger \mathrm{e}^{t Q^{\dagger} L^\dagger} Q^{\dagger} \rho(0)
    +\int_0^t \mathrm{d} s P^{\dagger} L^\dagger \mathrm{e}^{s Q^{\dagger} L^\dagger} Q^{\dagger} L^\dagger P^{\dagger} \rho(t-s) .%    \end{aligned}
\end{equation}
where $Q^\dagger = 1 - P^\dagger$ is the orthogonal projection, and $\rho(0) $ is the initial distribution. 
%Apart from using the pinned distribution $f_A(B)$ as the reference, the above procedure follows Zwanzig's projection-operator method, originally applied for equilibrium media.  

In the following, we apply the approximation from time-scale separation and only track the term up to $\epsilon^2$. 
Meanwhile, we concentrate on the long-time behavior with $t \sim O(\epsilon^{-2})$, which is the time-scale of the friction and noise effects as we will see later. 
The second term on the right-hand side in \eqref{exact} denotes the effect from the initial distribution. 
It vanishes if the initial distribution can be decomposed as the system distribution times the pinned distribution. 
For more general cases, for the long-time behavior, this term is negligible (usually decays exponentially with time $t$) as well.\\ 
The first term on the right-hand side in \eqref{exact} is first order in $\epsilon$, equal to
\begin{equation}
    \epsilon P^\dagger L_1^\dagger P^\dagger \rho(t) .
\end{equation}
The leading order of the last term in \eqref{exact} is $\epsilon^2$, and 
it brings forward the noise and the friction. 
To order $O(\epsilon^2)$, we can do the following approximation (noticing that $\rho(t-s)=\ue^{-sL^\dagger}\rho(t)$), 
\begin{equation}\label{appro}
    P^{\dagger} L^\dagger \mathrm{e}^{s Q^{\dagger} L^\dagger} Q^{\dagger} L^\dagger P^{\dagger} \mathrm{e}^{-s L^\dagger}\rho(t)=\epsilon^2 P^{\dagger} L_1^\dagger\left(\mathrm{e}^{s L_0^\dagger}-P^{\dagger}\right) L_1 ^\dagger P^{\dagger}\rho(t). 
\end{equation}
For the long-time behavior, $t\sim O(\epsilon^{-2})$, the Markov approximation is valid in the significant order. 
In this approximation, we set the upper boundary of the time-integral in \eqref{exact} to infinity, and the last term becomes 
\begin{equation}\label{Mark}
    \int_0^\infty \mathrm{d} s P^{\dagger}\epsilon L^\dagger_1 \left(\mathrm{e}^{s L^\dagger_0}-P^{\dagger}\right)  \epsilon L^\dagger_1 P^{\dagger} \rho(t) .
\end{equation}
We now show why Markov approximation is always valid. 
The error of the Markov approximation is 
\begin{equation}\label{error}
    \epsilon^2\int_t^\infty \mathrm{d} s P^{\dagger} L^\dagger_1 \left(\mathrm{e}^{s L^\dagger_0}-P^{\dagger}\right)  L^\dagger_1 P^{\dagger} \rho(t)= \epsilon^2\int_t^\infty \ud s \mathcal C(s) .
\end{equation}
As we will see in the final expression in \eqref{2nd} and \eqref{2nd2}, $\mathcal C(s)=\ev{\cdots}_\mathcal{A}$ can be finally expressed as a time-dependent correlation function in the medium dynamics only, which does not depend on $\epsilon$.  
As long as the time-integral is convergent, the error in \eqref{error} is an infinitesimal of higher order than $\epsilon^2$. 
Therefore, the Markov approximation for the probe dynamics will be exact in the limit $\epsilon \rightarrow 0$, $t\rightarrow \infty$, and $\epsilon^2 t=\mathrm{const}$. 

Eq.\eqref{exact} is for the total distribution $\rho$. 
We now take the integral over the media on both sides to obtain the equation for the reduced distribution $\tilde \rho=\tr_\mathcal{B}\rho=\int\ud \mathcal{B}\rho$, 
\begin{equation}\label{trace}
    \pdv{t}\tilde \rho(t)=\tr_\mathcal{B} \epsilon L^\dagger_1 P^\dagger \rho+ \int_0^\infty \mathrm{d} s \tr_\mathcal{B}\epsilon L^\dagger_1 \left(\mathrm{e}^{s L^\dagger_0}-P^{\dagger}\right)  \epsilon L^\dagger_1 P^{\dagger} \rho(t).
\end{equation}
We have used $\tr_\mathcal{B} P^\dagger = \tr_\mathcal{B}$. 
After applying the expression of $P^\dagger$, the first term reads 
\begin{equation}\label{induced}
    \begin{aligned}
    \int\ud \mathcal{B} \epsilon L_1^\dagger [f_\mathcal{A}(\mathcal{B}) \tilde \rho(\mathcal{A},t)]
    =-\int \ud \mathcal{B} \epsilon \sum_i L_1\mathcal{A}_i \pdv{\mathcal{A}_i} [\tilde \rho(\mathcal{A},t)]
    =-\sum_{i} \pdv{\mathcal{A}_i}[\ev{L\mathcal{A}_i}_{\mathcal{A}}\tilde \rho(\mathcal{A},t)] .
    \end{aligned}
\end{equation}
Here and in the following, $\ev{\cdots}_\mathcal{A}$ denotes the average in the pinned dynamics (Born-Oppenheimer quasistatic dynamics), and the superscript $\mathrm{bo}$ in the main text is omitted here for simplicity. 
In the first equality, we use $L_0 f_\mathcal{A}(\mathcal{B})=0$. 
In the second equality, we use that $L_1$ is conservative so that $L_1^\dagger =-L_1$, and we use the chain rule because it is a first order differential operator, 
In the third equality, we use that $\epsilon L_1\mathcal{A}_i=L\mathcal{A}_i$, and the divergence of $L\mathcal{A}_i$ is zero, $\sum_i\partial_{\mathcal{A}_i} L\mathcal{A}_i =0$. 
The final result gives the drift term of the induced force, which is to the first order of $\epsilon$. 
No approximation is made for this induced force.

The second term in \eqref{trace} reads 
\begin{equation}\label{2nd}
    \begin{aligned}
    &\int_0^\infty \ud s\int\ud \mathcal{B} \epsilon L_1^\dagger\left( \mathrm{e}^{s L_0^\dagger}-P^{\dagger}\right) \epsilon L_1^\dagger P^{\dagger} \rho(t)\\
    &=\int_0^\infty \ud s\int\ud \mathcal{B}^\prime\ud \mathcal{A}^\prime \delta(\mathcal{A}-\mathcal{A}^\prime)\epsilon L^{\dagger\prime}_1\left(\mathrm{e}^{s L_0^{\dagger\prime}}-P^{\dagger\prime}\right) \epsilon L^{\dagger\prime}_1 [f_{\mathcal{A}^\prime}(\mathcal{B}^\prime) \tilde\rho(\mathcal{A}^\prime,t)]\\
    &=\int_0^\infty \ud s\int\ud \mathcal{B}\ud \mathcal{A}^\prime (\ue^{s L^\prime_0}- P^\prime)\epsilon L_1^\prime[\delta(\mathcal{A}-\mathcal{A}^\prime)] \epsilon L^{\dagger\prime}_1 [f_{\mathcal{A}^\prime}(\mathcal{B}^\prime) \tilde\rho(\mathcal{A}^\prime, t)]\\
    &=\int_0^\infty \ud s\int\ud \mathcal{B}^\prime\ud \mathcal{A}^\prime \sum_{ij}  [(L^\prime \mathcal{A}^\prime(s) -P^\prime L^\prime \mathcal{A}^\prime)\pdv{\mathcal{A}_i}\delta(\mathcal{A}-\mathcal{A}^\prime)] [ L^\prime \mathcal{A}_j^\prime\pdv{\mathcal{A}^\prime_j}f_{\mathcal{A}^\prime}(\mathcal{B}^\prime)\tilde \rho(\mathcal{A}^\prime,t)]\\
    &= \sum_{i,j}\pdv{\mathcal{A}_i}\int_0^\infty \ud s\int\ud \mathcal{B}  (L \mathcal{A}_i(s) -\ev{L\mathcal{A}_i }_\mathcal{A})L \mathcal{A}_j  \pdv{\mathcal{A}_j}\left[f_\mathcal{A}(\mathcal{B}) \tilde \rho(\mathcal{A},t)\right] .
    \end{aligned}
\end{equation}
In the first equality, we plug in a delta function $\delta(\mathcal{A}-\mathcal{A}^\prime)$ and an integral over $\mathcal{A}^\prime$, and the integrated variable $\mathcal{B}$ is changed to $\mathcal{B}^\prime$. Meanwhile, the last projection $P^\dagger$ is written down explicitly. In the second equality, operator $\epsilon L^{\dagger\prime}_1\left(\mathrm{e}^{s L_0^{\dagger\prime}}-P^{\dagger\prime}\right)$ is moved to act on the delta function with its conjugate operator $(\ue^{s L^\prime_0}- P^\prime)\epsilon L_1^\prime$. 
In the third equality, we use the property of $L_1^\dagger$ again and replace $\ue^{s L^\prime_0} L\mathcal{A}^\prime$ with the time-evolution value in the pinned dynamics $L\mathcal{A}^\prime(s)$. 
In the fourth equality, $\partial_{\mathcal{A}_i}$ is moved to the front, 
and we do the integral of the delta function and change $\mathcal{B}^\prime$ back to $\mathcal{B}$. 
The final expression in the above equation can be split into two terms, 
\begin{equation}\label{2nd2}
    \sum_{i,j}\pdv{\mathcal{A}_i}\int_0^\infty \ud s\ev{\delta L\mathcal{A}_i(s)L\mathcal{A}_j\pdv{\mathcal{A}_j}\log f_\mathcal{A}(\mathcal{B}) }_\mathcal{A} \tilde \rho(\mathcal{A})
    +\sum_{i,j}\pdv{\mathcal{A}_i}\int_0^\infty \ud s\ev{\delta L\mathcal{A}_i(s)\delta L\mathcal{A}_j}_\mathcal{A} \pdv{\mathcal{A}_j}\tilde \rho(\mathcal{A}) ,
\end{equation}
with $\delta L\mathcal{A}=L\mathcal{A}-\ev{L\mathcal{A}}_\mathcal{A}$ the deviation from its average value. 
These two terms correspond to the friction and the noise effects, respectively. 
Since $L\mathcal{A}_i=\epsilon L_1\mathcal{A}_i$, they are to the order of $\epsilon^2$. 

In summary, the Fokker-Planck equation for the reduced distribution is 
\begin{equation}\label{FPE}
\begin{aligned}
    \pdv{t}\tilde \rho(\mathcal{A},t)&=-\sum_i\pdv{\mathcal{A}_i}\left[K_i(\mathcal{A}) \tilde \rho(\mathcal{A},t)\right]\\
    &-\sum_{i}\pdv{\mathcal{A}_i}\left[\mu_i(\mathcal{A}) \tilde \rho(\mathcal{A},t)\right]
    -\sum_{i}\pdv{\mathcal{A}_i}[\sum_j\pdv{D_{ij}(\mathcal{A})}{\mathcal{A}_j} \tilde \rho(\mathcal{A},t)]
    +\sum_{i,j}\pdv{\mathcal{A}_i}\pdv{\mathcal{A}_j}\left[D_{ij}(\mathcal{A}) \tilde \rho(\mathcal{A},t)\right] ,
    \end{aligned}
\end{equation}
with the following expressions, 
\begin{align}
    &K_i(\mathcal{A})=\ev{L\mathcal{A}_i}_{\mathcal{A}},
\\
    &\mu_i(\mathcal{A})=-\int_0^\infty \ud s\sum_j\ev{\delta L\mathcal{A}_i(s)L\mathcal{A}_j\pdv{\mathcal{A}_j}\log f_\mathcal{A} }_\mathcal{A} ,
\\
    &D_{ij}(\mathcal{A})=\ev{\delta L\mathcal{A}_i(s)\delta L\mathcal{A}_j}_\mathcal{A}  ,
\end{align}
which is Eq.~1 and Eq.~2 in the main text. 

\section{Nonequilibrium response theory and its measurable expression}
In this section, we present a derivation from Eq.~9 to Eq.~10 in the main text. 
We first introduce the general form of the nonequilibrium response theory \cite{over,frenesy,resp} and then use it to rewrite the friction coefficient to a measurable expression, Eq.~10 in the main text. 

We consider overdamped Brownian particles $x_i$ in contact with an environment at temperature $T$. 
The same final expression is also valid for Markov jump processes. 
The equation of motion is 
\begin{equation}
    \dot x_i = g_i(x) +\sqrt{\frac2\beta}\xi_i , 
\end{equation}
where $g_i$ are the forces on the particles, and $\xi_i$ are independent Gaussian white noise. 
We call the above process reference dynamics. 
Next, if there is a time-dependent perturbation on the above dynamics by adding a potential $-h(t)V(x)$ where $h(t)$ is a time-dependent small amplitude, the perturbed equation of motion will become 
\begin{equation}
    \dot x_i = g_i(x)+h(t)\pdv{x_i}V(x) +\sqrt{2T}\xi_i  .
\end{equation}
We suppose the perturbation starts at $t_0$, so $h(s)=0$ for $s<t_0$. 

From the path-integral approach, the probability of a trajectory $\omega$ (starting at $t_0$ and ending at $t_1$) conditioned on its initial value $\omega_0$ is expressed as 
\begin{equation}
    \Pr[\omega|\omega_0] \propto \exp(- \mathcal A_h[\omega]), 
\end{equation}
where the action $\mathcal A_h$ (should be distinguished with $\mathcal A$ in the previous section) is a stochastic integral under Stratonovich convention, 
\begin{equation}\label{action}
    \mathcal A_h = \int_{t_0}^{t_1}\ud s \sum_i \frac{1}{4T}\left[\dot x_i(s)-g_i(x(s))-h(s)\pdv{x_i}V(x(s))\right]^2 .
\end{equation}
When $h(s)=0$, the above action reduces to that of the reference dynamics, $\mathcal A_0$. 
Furthermore, that action can be divided into two parts according to the parity under time reversal, 
$
    \mathcal A_h=\frac 12 \mathcal S_h + \mathcal D_h, 
$
with the following expressions, 
\begin{align}\label{entropic}
    &\mathcal S_h = -\beta\int_{t_0}^{t_1}\ud s \sum_i \dot x_i \left[g_i(x(s))+h(s) \partial_iV(x(s))\right],\\
    &\mathcal D_h = \frac{\beta}{4}\int_{t_0}^{t_1}\ud s \sum_i\left[\dot x_i(s)^2+(g_i(x(s))+h(s)\partial_i V(x(s)))^2\right] . \label{frenetic}
\end{align}
Under time-reversal, the symmetric  $\mathcal D_h$ is the frenetic part, while the antisymmetric $\mathcal S_h$ is the entropic part. 
The entropic part is nothing but the trajectory-dependent change in entropy of the underlying thermal environment, \cite{local}.

With the path-probability, the average of an observable $F(x(t))$ at time $t$ ($t>t_0$) can be expressed as, 
\begin{equation}
\begin{aligned}
    \ev{F(t)}_h&=\int \ud \omega F(t) \exp(-\mathcal A_h[\omega]) \rho(\omega_0)= \int \ud \omega F(t)\exp(-\mathcal A_h[\omega]+\mathcal A_0[\omega]) \exp(-\mathcal A_0[\omega]) \rho(\omega_0)\\
    &= \ev{F(t)\exp(-\mathcal A_h[\omega]+\mathcal A_0[\omega])}_0 .
    \end{aligned}
\end{equation}
Here, $\ud \omega$ denotes the integral over all trajectories starting at $t_0$. $\ev{\cdots}_h$ and $\ev{\cdots}_0$ denote the average in the perturbed dynamics and reference dynamics, respectively. 
Therefore, to the first order of $h$, the difference of the average between the perturbed process and reference process is 
\begin{equation}
    \ev{F(t)}_h-\ev{F(t)}_0= \ev{F(t)[\exp(-\mathcal A_h+\mathcal A_0)-1]}_0 =-\int_{t_0}^t \ud s h(s)\ev{F(t)\frac{\delta}{\delta h(s)} \left(\frac 12 \mathcal S_h+\mathcal D_h\right)}_0 +O(h^2) .
\end{equation}
According to the expressions \eqref{entropic} \eqref{frenetic}, the above difference can be expressed as a correlation function only in the reference dynamics (to linear order): 
\begin{equation}
\begin{aligned}
    \ev{F(t)}_h-\ev{F(t)}_0=& \int_{t_0}^t\ud s  \frac{\beta }{2}h(s) \ev{F(t) \left[\sum_i \dot x_i(s)  \partial_i V(s) -\sum_i g_i(s)\partial_i V(s) \right]}_0 \\
    =& \int_{t_0}^t\ud s \frac{\beta}{2} h(s)\left[\dv{s}\ev{F(t) V(s)}_0-\ev{F(t) L_0 V(s) }_0 \right] .
    \end{aligned}
\end{equation}
We have used $\dv {s} V(s)=\sum_i \dot x_i \partial_i V$ and the backward generator of the reference dynamics, $L_0=\sum_i g_i\partial_i $. 
In the bracket of the last expression,  
two terms correspond to the contributions from the entropic term and the frenetic term, respectively. 
The above formula serves as a general result of response theory, a generalization of Green-Kubo formul{\ae} to nonequilibrium systems. 
A similar derivation based on path-probability yields the same final expression for discrete Markov jump processes. 

We now apply the above formula to Eq.~9 in the main text: 
\begin{equation}
    \gamma(s) \Delta q = \ev{F(q,\eta(s))}_{q+\Delta q\rightarrow q}^* - \ev{F(q,\eta(s))}_q .
\end{equation}
Here, the overdamped medium particles are denoted by $\eta$. 
The quantity to be averaged is the force $F(q,\eta(s))=-\partial_q U(q,\eta(s)) $ on the probe. 
The second term, the average in the pinned dynamics for medium with fixed $q$, is taken to be the reference. 
The perturbed dynamics is called star dynamics, where the medium undergoes a quench process. 
This process starts at the infinite past, $t_0=-\infty$. For $t<0$, the probe position is fixed at $q+\Delta q$ while for $t>0$, the probe position is $q$. 
In the star dynamics, medium particles satisfy the following equation of motion, 
\begin{equation}
\begin{aligned}
    \dot \eta_i &= g_i(\eta) -  \pdv{\eta_i} U(q+\theta(-t)\Delta q ,\eta) \\
    &=g_i(\eta) - \theta(-t)\Delta q \pdv{\eta_i} \pdv{q} U (q,\eta)  .
    \end{aligned}
\end{equation}
Here, $g_i(\eta)$ is the internal force between medium particles. We used the fact that $\Delta q$ is small in the second line. 
Compared to the general setup for the response theory, we choose the perturbation potential here to be $V=-\partial_q U = F$ and 
$h(s)=\theta (-s) \Delta q$. 
Using the general formula, we obtain
\begin{equation}
    \begin{aligned}
    \ev{F(s)}^*_{q+\Delta q\rightarrow q}-\ev{F(s)}_q &=\frac \beta 2 \int_{-\infty}^s\ud u \theta(-u)\Delta q[ \dv{u}\ev{F(s)F(u)}_q- \ev{F(s)L_0F(u) }_q]\\
    &=\Delta q \frac{\beta}2 \left[\ev{F(s)F(u)}_q\eval_{u=-\infty}^{0}-\int_{-\infty}^0\ud u\ev{L_0F(u),F(s)}_q \right]\\
    &=\Delta q \frac{\beta}2 \left[\ev{F(s),F(0)}_q-\int_s^\infty\ud u\ev{L_0F(0),F(u)}_q \right] .
    \end{aligned}
\end{equation}
This accounts for Eq.~10 in the main text. 
The friction coefficient $\gamma=\int_0^\infty \ud s \gamma(s)$ is now divided into two terms, 
\begin{align}
    \gamma_{\text{ent}}&= \frac{\beta}{2} \int_0^\infty\ud s  \ev{F(s),F(0)}_q\\
    \gamma_{\text{fre}}&= -\frac{ \beta}2  \int_0^\infty\ud s \int_s^\infty\ud u\ev{L_0F(0),F(u)}_q .
\end{align}
They are the entropic and frenetic parts of the friction coeffcient, respectively. 

\section{Run-and-tumble particles on a ring} 
We consider the run-and-tumble medium model defined by the Eqs. 12--13 in the main text. 
We recall that we need to compute the friction
\begin{equation}\label{fri}
\gamma(s) = \ev{Lp(s) \pdv{q}\log f_q(x,\sigma) }_q^\text{bo}
\end{equation}
as for Eq.~6 in the main text. That requires computing the pinned distribution $f_q$, and for the time-correlation, to compute the transition probabilities.\\
$x$ and $q$ are the positions of the RTP particle and the heavy probe on the ring, respectively.
For fixed particle position $q$, we restrict $x-q$ to the interval $(-\pi,\pi]$ and we
apply the periodic boundary condition. 
In the following, we consider the case $q=0$ and solve the dynamics of the RTP-particle.

\subsection{Eigenvalue problem for the master equation}
The Master equation for the run-and-tumble particles is 
\begin{equation}
    \pdv{t} f(x,\sigma,t) = -\pdv{x} [(F_m+v\sigma) f(x,\sigma,t)] +\alpha [f(x,-\sigma,t)-f(x,\sigma,t)] ,
\end{equation}
with connection conditions at $x=0$ and $x=\pi$: 
\begin{equation}\label{connection}
    \begin{aligned}
    &[(v\sigma-k) f(x,\sigma)]_{x=0+} = [(v\sigma+k) f(x,\sigma)]_{x=0-} \\
    &[(v\sigma-k) f(x,\sigma)]_{x=+\pi} = [(v\sigma+k) f(x,\sigma)]_{x=-\pi} .
    \end{aligned}
\end{equation}
The eigenvalue problem reads
\begin{equation}
-\pdv{x} [(F_m+v\sigma) f(x,\sigma)] +\alpha [f(x,-\sigma)-f(x,\sigma)] =\lambda f(x,\sigma) .
\end{equation}
and we find eigenvalues  
\begin{equation}
    \lambda=0, \lambda=-2\alpha, \frac{L \sqrt{\alpha ^2 k^2+\lambda  v^2 (2 \alpha +\lambda )}}{v^2-k^2}=n \pi \ui, n\in \mathbb{N}_+ .
\end{equation}
The solution to the last equation gives 
\begin{equation}
    \lambda_n^\pm = -\alpha \pm \frac{\sqrt{\alpha^2 (v^2-k^2)- n^2\pi^2(v^2-k^2)^2 /L^2}}{v}, \quad n\in \mathbb{N}_+ .
\end{equation}
We denote $b_n=\sqrt{ n^2\pi^2 /L^2-\alpha^2 /(v^2-k^2)}$, so that $\lambda_n^\pm =-\alpha \pm \ui b_n(v^2-k^2)/v $.
There is only one zero eigenvalue, and all other eigenvalues have negative real parts, 
meaning that the steady-state distribution is unique. 
For $\lambda=0$, the eigenfunction is also the steady-state distribution: 
\begin{equation}
    f_0(x,\sigma)\propto (v+\sigma k\operatorname{sgn}x )\exp\left(-\frac{2k \alpha}{v^2-k^2}\abs{x}\right)  .
\end{equation}
For $\lambda=-2\alpha$, the eigenfunction is
\begin{equation}
    f_{-2\alpha}(x,\sigma)\propto (\sigma v+k\operatorname{sgn}x)\exp\left(\frac{2k \alpha}{v^2-k^2}\abs{x}\right) .
\end{equation}
For $\lambda_n^-$, the eigenfunction space is two-fold degenerated, represented by two arbitrary coefficient $\mathcal A_\pm$ (should be distinguished with $\mathcal A$ in the previous two sections), 
\begin{equation}
    \begin{aligned}
   &f_{\lambda_n^-}(x,\sigma)=
   \exp\left(\ui b_n\frac{k}{v} \abs{x} \right)+ [\mathcal A_\sigma \left(v+\sigma k\operatorname{sgn}x\right) \cos (nx)\\
   & +\sin (n x)
   \left(\mathcal A_\sigma\frac{\sigma v+k\operatorname{sgn}x}{n}\ui b_n+ \mathcal A_{-\sigma} \sigma \frac{\alpha}{n}
   \right)] .
    \end{aligned}
\end{equation}
For $\lambda_n^+$, the eigenfunction is also two-fold degenerated, 
\begin{equation}
    \begin{aligned}
   &f_{\lambda_n^+}(x,+1)=
   \exp\left(-\ui b_n\frac{k}{v} \abs{x} \right)   [\mathcal A_\sigma \left(v+\sigma k\operatorname{sgn}x\right) \cos (nx)\\
   & +\sin (n x)
   \left(\mathcal A_\sigma\frac{-\sigma v-k\operatorname{sgn}x}{n}\ui b_n+\mathcal A_{-\sigma}  \sigma\frac{\alpha}{n}
   \right)] ,
    \end{aligned}
\end{equation}
where $\mathcal A_\sigma$ are arbitrary coefficients. 
We choose the linear-independent basis $f_{\lambda_n^\pm}^{1,2}$ in the eigenfunction space by setting $\mathcal A_{1}=1,\mathcal A_{-1}=0$ and $\mathcal A_{1}=0,\mathcal A_{-1}=1$. 

\subsection{Orthogonality of the eigenfunctions}
Next, we consider the eigenvalue problem for the backward generator $L$, which gives left eigenfunctions of $L^\dagger$, 
\begin{equation}
    Lg(x,\sigma) =[F_m(x)+v\sigma]\pdv{x} g(x,\sigma) +\alpha [g(x,-\sigma)-g(x,\sigma)] .
\end{equation}
This time, the connection condition is just the continuity of the eigenfunction at $x=0$ and $x=L$,
\begin{equation}
    \begin{aligned}
    &[g(x,\sigma)]_{x=0+} = [g(x,\sigma)]_{x=0-} \\
    &[g(x,\sigma)]_{x=+L} = [g(x,\sigma)]_{x=-\pi} .
    \end{aligned}
\end{equation}
The eigenvalues are the same as before, $\lambda=0,-2\alpha,\lambda_n^\pm$. 
For $\lambda=0$, the eigenfunction is
\begin{equation}
    g_0(x,\sigma)\propto 1, \quad x\in (-\pi,\pi) .
\end{equation}
For $\lambda=-2\alpha$, the eigenfunctions are
\begin{equation}
    g_{-2\alpha}(x,\sigma)\propto \sigma, \quad x\in (-\pi,\pi) .
\end{equation}
For $\lambda=\lambda_n^-$, there are also two fold degeneracy with eigenfunction 
\begin{equation}
    g_{\lambda_n^-}(x,\sigma)= \exp(-\ui \frac{k}{v} b_n \abs {x})
    [\mathcal B_\sigma \cos(n x) + ( -\mathcal B_{-\sigma} \frac{\alpha}{n(\sigma v- k\operatorname{sgn} x)} -\sigma\mathcal B_{\sigma} \ui b_n/n )\sin(nx)] .
\end{equation}
For $\lambda=\lambda_n^+$, the eigenfunction is 
\begin{equation}
    g_{\lambda_n^+}(x,\sigma)= \exp(\ui \frac{k}{v} b_n \abs {x})
    [\mathcal B_\sigma \cos(n x) + (- \mathcal B_{-\sigma} \frac{\alpha}{n(\sigma v- k\operatorname{sgn} x)} +\sigma \mathcal B_\sigma \ui b_n/n )\sin(nx)] .
\end{equation}
We choose the basis $f_{\lambda_n^\pm}^{1,2}$ by setting $\mathcal B_{1}=1,\mathcal B_{-1}=0$ and $\mathcal B_{1}=0,\mathcal B_{-1}=1$.

The left and right eigenfunctions with different eigenvalues are orthogonal, $(g_\mu,f_{\lambda})=0,  \mu\neq \lambda$, with respect to the inner product 
\begin{equation}
    \langle f,g\rangle = \int_{-L}^L\ud x \sum_\sigma f(x,\sigma) g(x,\sigma) .
\end{equation}
The inner products of eigenfunctions with the same eigenvalue are 
\begin{equation}
    (g_0,f_0)=\frac{2}{k\alpha}v(v^2-k^2)[1-\exp(\frac{-2kL\alpha}{v^2-k^2})],\quad    (g_{-2\alpha},f_{-2\alpha})=\frac{2}{k\alpha}v(v^2-k^2)[\exp(\frac{2kL\alpha}{v^2-k^2})-1] ,
\end{equation}
and for $\lambda_n^\pm$, 
\begin{equation}
    (g_{\lambda_n^+}^i,f_{\lambda_n^+}^j)=M^{ij}_{\lambda_n^+},\qquad
    (g_{\lambda_n^-}^i,f_{\lambda_n^-}^j)=M^{ij}_{\lambda_n^-} ,
\end{equation}
where 
\begin{equation}
    M^{ij}_{\lambda_n^-}=
    \begin{pmatrix}
        \frac{2\pi vb_n^2}{n^2} & 
        -\frac{2\pi v^2\alpha\ui b_n}{n^2(v^2-k^2)}\\
        -\frac{2\pi v^2\alpha\ui b_n}{n^2(v^2-k^2)} &
        \frac{2\pi vb_n^2}{n^2}
    \end{pmatrix},\;
    M^{ij}_{\lambda_n^-}=
    \begin{pmatrix}
        \frac{2\pi vb_n^2}{n^2} & 
        \frac{2\pi v^2\alpha\ui b_n}{n^2(v^2-k^2)}\\
        \frac{2\pi v^2\alpha\ui b_n}{n^2(v^2-k^2)} &
        \frac{2\pi vb_n^2}{n^2}
    \end{pmatrix} .
\end{equation}

\subsection{Steady-state distribution and transition probability}
For a generic fixed value of particle position $q$, 
the steady-state distribution for the run-and-tumble particle is 
obtained by replacing $x$ with $x-q$ in the expression of the steady-state distribution,
\begin{equation}\label{ss}
    f_q(x,\sigma)=(v+\sigma k\operatorname{sgn}(x-q) )\exp\left(-\frac{2k \alpha}{v^2-k^2}\abs{x-q}\right)  /Z_0 .
\end{equation}
According to the expression of the left and right eigenfunctions, the normalization coefficient is $Z_0=(g_0,f_0)$.

The transition probability, $P(x,\sigma,t|x_0,\sigma_0)$ is the solution to the Fokker-Planck equation with the initial distribution $\delta(x-x_0)\delta_{\sigma,\sigma_0}$. 
The delta-function can be expanded in terms of the eigenfunctions,
\begin{equation}
    \delta(x-y)\delta_{\sigma,\sigma'} = \sum_\lambda f_\lambda(x,\sigma) d_\lambda(y,\sigma') .
\end{equation}
According to the orthogonality of the eigenfunctions, 
\begin{equation}
%    \begin{aligned}
    d_\lambda(y,\sigma^\prime) = \sum_\mu M^{-1}_{\lambda \mu}(g_\mu(x,\sigma),\delta(x-y)\delta_{\sigma,\sigma'})
    =\sum_\mu M^{-1}_{\lambda \mu}g_\mu(y,\sigma^\prime) .
 %   \end{aligned}
\end{equation}
The transition probability is expressed as 
\begin{equation}\label{transition_probability}
   % \begin{aligned}
    P(x,\sigma,t|x_0,\sigma_0)= \sum_\lambda \exp(\lambda t) f_\lambda(x,\sigma) d_\lambda(x_0,\sigma_0) =\sum_{\lambda\nu} \exp(\lambda t) f_\lambda(x,\sigma) M^{-1}_{\lambda \nu}g_\nu(x_0,\sigma_0) .
   % \end{aligned}
\end{equation}

\section{Friction from the RTP-medium on the ring}
From the expression \eqref{ss} for the steady-state distribution, the friction coefficient \eqref{fri} can be expressed as
\begin{equation}
    \gamma(s) = \ev{ k\operatorname{sgn}(x(s)) \left[\frac{2k\alpha}{v^2-k^2}\operatorname{sgn}x-\frac{2\sigma k}{v}\delta(x)+\frac{2\sigma k }{v}\delta(x-L)\right]}_{q=0} ,
\end{equation}
which can be splitted into two parts,
\begin{equation}\label{g1_def}
    \gamma_1(s)=\frac{2k^2\alpha}{v^2-k^2} \ev{\operatorname{sgn}(x(s))\operatorname{sgn}x}_{q=0} 
\end{equation}
and 
\begin{equation}\label{g2_def}
    \gamma_2(s)=\frac{2k^2}{v} [\ev{-\operatorname{sgn}(x(s))\sigma\delta(x)}_{q=0}+\ev{\operatorname{sgn}(x(s))\sigma\delta(x-L)}_{q=0}] .
\end{equation}
$\gamma_1 = 2\gamma_\text{ent}$ is the ``usual'' friction coefficient, which is proportional to the variance of the noise and is always positive;
$\gamma_2 = \gamma_\text{fren}-\gamma_\text{ent}$ is always negative, as will follow. 
In the following, we omit the subscript $q=0$ for simplicity. 

We first calculate $\gamma_2(s)$. 
It turns out that 
\begin{equation}\label{gamma2_relation}
    \gamma_2(s)=\frac{k^2}{v^2-k^2}\dv{s} \ev{\operatorname{sgn}x(s)\operatorname{sgn} x(0)} .
\end{equation}
To prove the above relation, we rewrite the right-hand side 
\begin{equation}
    \begin{aligned}
    &\dv{s}\ev{\operatorname{sgn}x(s)\operatorname{sgn} x(0)}\\
    &=\int\ud x_s \int\ud x_0\sum_{\sigma_s}\sum_{\sigma_0} \pdv{s} P(x_s,\sigma_s,s|x_0,\sigma_0) f_0(x_0,\sigma_0)\operatorname{sgn}x_s \operatorname{sgn}x_0\\
    &=\int\ud x_s \int\ud x_0\sum_{\sigma_s}\sum_{\sigma_0} L_{0} P(x_s,\sigma_s,s|x_0,\sigma_0) f_0(x_0,\sigma_0)\operatorname{sgn}x_s \operatorname{sgn}x_0\\
    &=\int\ud x_s \int\ud x_0\sum_{\sigma_s}\sum_{\sigma_0} P(x_s,\sigma_s,s|x_0,\sigma_0) \operatorname{sgn}x_s L_{0}^\dagger[ f_0(x_0,\sigma_0) \operatorname{sgn}x_0] .
    \end{aligned}
\end{equation}
Plugging in the expression of the forward operator and using $L^\dagger f_0=0$, we find the formula in the above equation can be written as 
\begin{equation}
    \int\ud x_s \int\ud x_0\sum_{\sigma_s}\sum_{\sigma_0} P(x_s,\sigma_s,s|x_0,\sigma_0) \operatorname{sgn}x_s  f_0(x_0,\sigma_0) 2(v\sigma-\operatorname{sgn}x_0)(-\delta(x) + \delta(x-\pi)) .
\end{equation}
Meanwhile, the left-hand side of \eqref{gamma2_relation} is 
\begin{equation}
    \begin{aligned}
    &\frac{2k^2}{v} [\ev{-\operatorname{sgn}(x(s))\sigma\delta(x)}_{q=0}+\ev{\operatorname{sgn}(x(s))\sigma\delta(x-L)}_{q=0}]\\
    &=\int\ud x_s \int\ud x_0\sum_{\sigma_s}\sum_{\sigma_0} P(x_s,\sigma_s,s|x_0,\sigma_0) \operatorname{sgn}x_s  f_0(x_0,\sigma_0) \frac{2k^2}{v+\sigma k\operatorname{sgn}x_0}(-\delta(x) + \delta(x-\pi)) .
    \end{aligned}
\end{equation}
From the steady-state distribution \eqref{ss}, we find 
\begin{equation}
    \int\ud x_0 f_0(x_0,\sigma_0) \frac{2k^2}{v+\sigma k\operatorname{sgn}x_0}(-\delta(x) + \delta(x-\pi))=\frac{k^2}{v^2-k^2}\int\ud x_0 f_0(x_0,\sigma_0) 2[v\sigma-\operatorname{sgn}x_0](-\delta(x) + \delta(x-\pi)) .
\end{equation}
Combining the above three equations explains \eqref{gamma2_relation}. 
Therefore, the time-integrated $\gamma_2$ is 
\begin{equation}
\begin{aligned}
    \int_0^{+\infty}\ud s\, \gamma_2(s)&=\frac{k^2}{v^2-k^2}\dv{s} \left[\ev{\operatorname{sgn}x(+\infty)\operatorname{sgn} x(0)}-\ev{\operatorname{sgn}x(0)\operatorname{sgn} x(0)}\right]\\
    &=-\frac{k^2}{v^2-k^2} ,
    \end{aligned}
\end{equation}
which is always negative and independent of $\alpha$. 

Next, we calculate $\gamma_1(s)$. 
The time-integrated coefficient can be expressed via the transition probability,
\begin{equation}
    \begin{aligned}
    &\gamma_1=\int_0^\infty\ud s\,\frac{2k^2\alpha}{v^2-k^2} \ev{\operatorname{sgn}(x(s))\operatorname{sgn}x}\\
    &=\frac{2k^2\alpha}{v^2-k^2} \int\ud x_s \int\ud x_0\sum_{\sigma_s}\sum_{\sigma_0} P(x_s,\sigma_s,s|x_0,\sigma_0) f_0(x_0,\sigma_0)\operatorname{sgn}x_s \operatorname{sgn}x_0 .
    \end{aligned}
\end{equation}
Substituting the series expression \eqref{transition_probability} of the transition probability, and performing the integration, we get
\begin{equation}
    \begin{aligned}
    &\gamma_1=\frac{k^4}{v^2(v^2-k^2)}+
    \sum_{n=1}^{+\infty} \frac {8 n^2 vk^3\alpha^2(v^2-k^2) }{\pi  \left(\alpha ^2 k^2+n^2 \left(v^2-k^2\right)^2\right)^3 }\times\\
    &\{-(-1)^n \left(n^2 \left(v^2-k^2\right)^2-\alpha ^2 \left(2v^2-k^2\right)\right) 
    \frac{1}{b_n}\sin \left(\frac{\pi  k b_n}{v}\right)\\
    &+2 \alpha  v (v^2-k^2)\left[1-(-1)^n 
    \cos \left(\frac{\pi  k b_n}{v}\right)\right]\coth(\frac{\pi k\alpha}{v^2-k^2})\} ,
    \end{aligned}
\end{equation}
where $b_n=\sqrt{ n^2-\alpha^2 /(v^2-k^2)}$. 
This expression involves the sum of the series, which converges fast. \\

Therefore, $\gamma_1$ only depends on $\frac{v}{k}$ and $\frac{\alpha}{k}$
and $\gamma_2$ only depends on $\frac{v}{k}$, and the total friction coefficient can be written as $\gamma(v/k,\alpha/k)$. 
For the limit of $\alpha\rightarrow0$ (but still larger than the relaxation rate of the probe), 
which corresponds to infinite persistence, 
 the total friction coefficient becomes $\gamma=-\frac{k^2}{v^2} $.

For the noise amplitude, we have 
\begin{equation}
    B=\int_0^\infty\ud s\ev{k\operatorname{sgn}(x(s)) ,k\operatorname{sgn}x}_{q=0}.
\end{equation}
Comparing it with the entropic part of the friction coefficient, $\gamma_{\mathrm{ent}}=\gamma_1/2$ and Eq. \eqref{g1_def}, we identify the effective inverse temperature for the medium as $\beta_{\mathrm{eff}}=\frac{2\alpha}{v^2-k^2}$ .

\bibliography{reduced.bib}